\documentclass[12pt]{iopart}
\usepackage[numbers,square,sort&compress]{natbib}

\usepackage[T1]{fontenc} % output font encoding with special characters
\usepackage[utf8]{inputenc} % input font encoding with special characters
\usepackage{mathptmx} % Times New Roman-lke font (Nimbus Roman No9 L)
\usepackage{graphicx}
\usepackage{enumerate}
\usepackage[nolist]{acronym}
\usepackage{pdflscape}
\usepackage{afterpage}

\makeatletter
\@namedef{ver@amsmath.sty}{}
\makeatother
\usepackage{amstext}
\usepackage[version=3]{mhchem}
\usepackage{hyphenat}
\usepackage{xcolor}

\newcommand{\revision}[1]{#1}

\graphicspath{{./images/}}

\begin{document}

%\title[Interfaces in organic semiconductors]{The interplay of interfaces, supramolecular assembly, and electronics in organic semiconductors}
\topical[Interfaces in organic semiconductors]{The interplay of interfaces, supramolecular assembly, and electronics in organic semiconductors}

\author{Belinda J Boehm, Huong TL Nguyen and David M Huang}

\address{Department of Chemistry, School of Physical Sciences, The University of Adelaide, SA 5005, Australia}
\ead{david.huang@adelaide.edu.au}
\vspace{10pt}
%\begin{indented}
%\item[]April 2019
%\end{indented}

\begin{abstract}
Organic semiconductors, which include a diverse range of carbon-based small  molecules  and  polymers  with  interesting  optoelectronic  properties,  offer  many advantages over conventional inorganic semiconductors such as silicon and are growing in  importance  in  electronic applications. Although  these  materials  are  now  the basis  of  a  lucrative  industry in  electronic  displays,  many  promising  applications such  as  photovoltaics  remain largely  untapped. One  major  impediment  to  more rapid  development  and  widespread adoption  of  organic  semiconductor  technologies is  that  device  performance  is  not easily  predicted  from  the  chemical  structure  of the  constituent  molecules. Fundamentally,  this  is  because  organic  semiconductor molecules, unlike inorganic materials, interact by weak non-covalent forces, resulting in  significant  structural  disorder  that  can  strongly  impact  electronic  properties. Nevertheless,  directional forces  between  generally  anisotropic  organic-semiconductor molecules,  combined  with  translational symmetry  breaking  at  interfaces,  can  be  exploited  to control supramolecular order and consequent electronic properties in these materials. This review surveys recent advances in understanding of supramolecular assembly at organic-semiconductor  interfaces  and  its  impact  on  device  properties  in  a  number of  applications,  including  transistors,  light-emitting  diodes,  and  photovoltaics.  Recent  progress  and  challenges  in  computer  simulations  of supramolecular assembly and orientational anisotropy at these interfaces is also addressed.
\end{abstract}

\vspace{2pc}
\maketitle

%%%%%%%%%%%%%%%%%%%%%%%%%%%%%%%%%%%%%%%%%%%%%%%%%%%%%%%%%%

%----------------------------------------------------
%   ACRONYMS
%----------------------------------------------------

%------------------------
% DEVICES
%------------------------
\begin{acronym}
\acro{OLED}{organic light-emitting diode}
\acro{OFET}{organic field effect transistor}
\acro{BHJ}{bulk heterojunction}
\acro{FET}{field-effect transistor}
\acro{OPV}{organic photovoltaic}
\acro{OSC}{organic solar cell}
\acro{TFT}{thin-film transistor}
\acro{Voc}[$V_\mathrm{OC}$]{open-circuit voltage}
\end{acronym}

%------------------------
% SIMULATION METHODS
%------------------------
\begin{acronym}
\acro{MD}{molecular dynamics}
\acro{CG}{coarse-grained}
\acro{CG MD}{coarse-grained molecular dynamics}
\acro{LJ}{Lennard-Jones}
\acro{MC}{Monte Carlo}
\acro{DFT}{density functional theory}
\acro{IBI}{iterative Boltzmann inversion}
\acro{FM}{force matching}
\acro{MS CG}{multiscale coarse-graining}
\acro{GB}{Gay--Berne}
\acro{KMC}{kinetic Monte Carlo}
\acro{FG}{fine-grained}
\acro{GC}{grand canonical}
\end{acronym}

%------------------------
% POLYMERS/MOLECULES
%------------------------
\begin{acronym}
\acro{PBTTT}{poly(2,5\hyp{}bis(3\hyp{}alkylthiophen\hyp{}2\hyp{}yl)thieno[3,2\hyp{}bithiophene)}
\acro{P3HT}{poly(3-hexylthiophene-2,5-diyl)}
\acro{PBDTTPD}{poly-benzo[1,2-b:4,5-b']dithiophene-thieno[3,4-c]pyrrole-4,6-dione}
\acro{PC61BM}[PC$_{61}$BM]{[6,6]-phenyl-C$_{61}$-butyric acid methyl ester}
\acro{N2200}[P(NDI2OD-T2)]{poly{[N,N'\hyp{}bis(2\hyp{}octyldodecyl)naphthalene\hyp{}1,4,5,8\hyp{}bis(dicarboximide)\hyp{}2,6\hyp{}diyl]\hyp{}alt-5,5'\hyp{}(2,2'\hyp{}bithiophene)}}
\acro{TPD}{N,N'-bis(3-methylphenyl)-N,N'-diphenylbenzidine}
\acro{NDI}{naphthalene diimide}
\acro{SMDPPEH}{2,5-di(2-ethylhexyl)-3,6-bis(5''-\textit{n}-hexy-[2,2',5',2'']terthiophen-5-yl)-pyrrolo[3,4-c]pyrrolo-1,4-dione}
\acro{PDI}{perylene diimide}
\acro{P3AT}{poly(3-alkylthiophene)}
\acro{CBP}{4,4′-bis(N-carbazolyl)biphenyl}
\end{acronym}

%------------------------
% OTHER
%------------------------
\begin{acronym}
\acro{Tsub}[$T_\mathrm{sub}$]{substrate temperature}
\acro{Tg}[$T_\mathrm{g}$]{glass transition temperature}
\acro{GSP}{giant surface potential}
\acro{SAM}{self-assembled monolayer}
\acro{RDF}{radial distribution function}
\acro{MW}{molecular weight}
\acro{vdw}[vdW]{van der Waals}
\acro{CTS}{charge transfer state}
\acro{CT}{charge transfer}
\end{acronym}

%%%%%%%%%%%%%%%%%%%%%%%%%%%%%%%%%%%%%%%%%%%%%%%%%%%%%%%%%%%%%%%%%%

%------------------------------------------------------
%    INTRODUCTION
%------------------------------------------------------

\section{Introduction}
\label{Section: Introduction}

Semiconductors underpin much of the technology that drives modern society, from computers to mobile phones and solar panels. Although this technology is dominated by inorganic materials such as silicon, organic semiconductors comprising carbon-based molecules and polymers are rapidly growing in importance, and are now the basis of a lucrative industry, particularly for electronic device displays. Organic semiconductors offer many advantages over conventional inorganic materials: they can be processed from solution by energy-efficient methods and only thin films are needed to produce functional devices, saving on material requirements and costs, and making them compatible with high-throughput printing processes and flexible substrates \cite{Arias2010}. They can also potentially enhance inorganic semiconductor device performance, e.g. as singlet-fission or triplet-annihilation layers to improve light harvesting in silicon solar cells \cite{Tayebjee2015}. Nevertheless, many promising applications of organic semiconductors remain largely untapped, such as photovoltaics and energy-efficient lighting.

One of the main impediments to more rapid development and widespread adoption of organic semiconductor technologies is that device performance cannot easily be predicted from the chemical structure of the constituent molecules. Fundamentally, this is because molecules in organic semiconductor materials are held together by relatively weak non-covalent forces, resulting in significant structural disorder that can impact device properties in ways that are hard to predict. In addition, organic semiconductor thin films are generally formed by deposition onto a substrate from the vapor or solution phase in a nonequilibrium process that is challenging to describe by simple theory. Inorganic semiconductors, on the other hand, generally have ordered crystal structures that simplify prediction of electronic properties. Consequently, organic semiconductor device optimization often proceeds by trial-and-error, with material properties and processing conditions varied until a desired result is achieved. Thus, given the many applications of organic semiconductors, a more predictive approach will have major commercial and societal impacts.

\begin{figure}[hbt]
\centering
\includegraphics[width = 8cm]{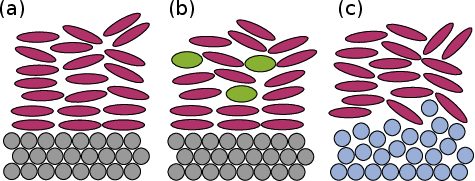}
\caption{Schematic of general organic semiconductor interfaces addressed in this review: (a) semiconductor-dielectric interface for \acp{OFET}; (b) semiconductor host-guest mixture on a substrate for \acp{OLED}; and (c) bulk-heterojunction electron donor-acceptor interface for \acp{OPV}.}
\label{Fig: generic interface schematic}
\end{figure}

One general property of almost all organic semiconductor molecules is the significant anisotropy in their shape and thus in their interactions. In principle, this molecular anisotropy can provide a handle by which structural order in organic semiconductors can be controlled. In particular, translational symmetry breaking at an interface can induce preferential orientation of anisotropic molecules \cite{DeGennes1995}. Ordering (and disorder) of organic semiconductor molecules at interfaces is known to have a large effect on electronic device performance in diverse applications and it is therefore important to understand the numerous mechanisms for its control. For example, aligning molecules with a permanent dipole moment at the interface between photo-active layer and electron contact in an organic solar cell can improve the power conversion efficiency by making the contact more selective to electrons over holes \cite{Wurfel2016}. The alignment of organic semiconductor molecules with respect to the dielectric surface in \acp{OFET} strongly impacts charge-carrier mobility \cite{Zhang2011a}. In \acp{OLED}, preferential alignment of host and guest emitter molecules with respect to the substrate surface can significantly boost the out-coupling efficiency of light by aligning guest transition dipole moments \cite{Kim2014}. In \acp{OPV}, significant variations in device efficiency have been correlated with molecular alignment at the interfaces between nanoscale domains of electron donor and acceptor materials \cite{Tumbleston2014}. A schematic representation of these interfaces is shown in figure \ref{Fig: generic interface schematic}.

Many examples, both experimental (e.g. \cite{Tumbleston2014, Lee2017, Ye2015, Osaka2015}) and computational (e.g. \cite{Yoneya2017a, Bagchi2019, Muccioli2011, Youn2018, Walters2017}), of semiconducting molecules displaying preferential alignment at interfaces can be found in the literature, but few general rules for predicting interfacial orientation from chemical structure or processing conditions appear to exist. Many factors have been implicated in control of interfacial orientation, including, among others, molecular shape \cite{Youn2018, Chen2013a, Dalal2015, Senes2017},  the presence, length, and composition of side chains \cite{Osaka2015, Lee2017, Ye2015, Osaka2014}, backbone planarity \cite{Chen2013},  temperature \cite{Dalal2015, Youn2018, Walters2017}, and solvent choice \cite{Youn2018, Chen2013, Luzio2013} and understanding the reasons behind these factors is important for the rational design of high performing organic semiconductors.

Although these interfaces are clearly important for understanding electronic processes in organic semiconductor-based devices, they are generally buried, making them experimentally difficult to characterize \cite{Salleo2007, Salleo2010}. While bulk morphology is relatively simple to characterize, morphology at the interface has been shown to often be different to that in the bulk, with changes in bulk morphology not necessarily correlated with that at the interface, and vice versa \cite{Kline2006, Duong2012, Hammond2011}. Although microstructural information can be gleaned from experimental techniques such as NEXAFS \cite{Nahid2017}, X-ray diffraction rocking \cite{Kline2006}, GIXD \cite{Duong2012}, and soft X-ray scattering techniques such as R-SoXS and P-SoXS \cite{Jiao2017}, molecular simulation can provide greater detail of the structures and processes occurring at these interfaces, with the advantage that it can often be directly coupled to an analysis of electronic structure and exciton and charge-carrier dynamics.

The use of molecular simulation to study organic-semiconductor interfaces has been recently reviewed \cite{Muccioli2014}, so it is not our goal to analyze in great depth the methods commonly used to study these systems, nor the general morphology at the interfaces. Instead we aim to provide an overview of the general role of molecular anisotropy at these interfaces and how it can be controlled to improve device performance through an understanding of the general physical principles that affect the interfacial microstructure. Additionally, we touch on the links between physical structure at the interfaces and electronic structure and charge transport characteristics to better understand the factors that improve or reduce device performance. 

\revision{We also do not specifically focus on organic semiconductor single crystals, which have been extensively and comprehensively reviewed previously \cite{Liu2009b,Virkar2010,Wang2012e,Dong2013,Li2016g,Zhang2018h,Wang2018k,Wang2019b}. Studies of single crystals have provided much important insight into the control of the bulk morphology of single-component small-molecule organic-semiconductor systems and the interfacial morphology of semiconductor--substrate interfaces with matching lattices, in which orientation with respect to the substrate is controlled by matching lattice vectors of a particular semiconductor crystal plane with those of the substrate surface, resulting in aligned epitaxial growth \cite{Zhang2018h}. However, such findings are not generally applicable to understanding the principles that govern interfacial microstructure and orientational ordering across different organic--semiconductor materials and device types, which are often significantly disordered and/or consist of multiple components. Even for a single crystal, the interface with a substrate is generally expected to be disordered and may even have a different structure from the bulk due to mismatched lattices and semiconductor--substrate interactions. Thus, such interfaces are expected and in some cases have been shown to be governed the same principles and processes discussed below for general organic-semiconductor systems with disorder.}

The outline of this review is as follows. We initially describe the theoretical background relevant to the ordering of anisotropic molecules at interfaces in section~\ref{Section: Theoretical background} before examining the role of these interfaces and the effects of molecular alignment at these interfaces in \acp{OFET}, \acp{OLED}, and \ac{BHJ} \acp{OPV} in section \ref{Section: Role of interfaces}. In section \ref{Section: General principles} we survey the available literature on interfacial orientational ordering of organic semiconductors with the aim of identifying general physical principles for predicting interfacial orientation. Finally, in section~\ref{Section: Modeling interfaces}, we discuss computer simulation methods for improving understanding of the structure and supramolecular assembly of organic-semiconductor interfaces, along with the challenges associated with accurately capturing experimentally relevant length and time scales in these simulations. 

\begin{figure*}
    \centering
    \includegraphics[width=14cm]{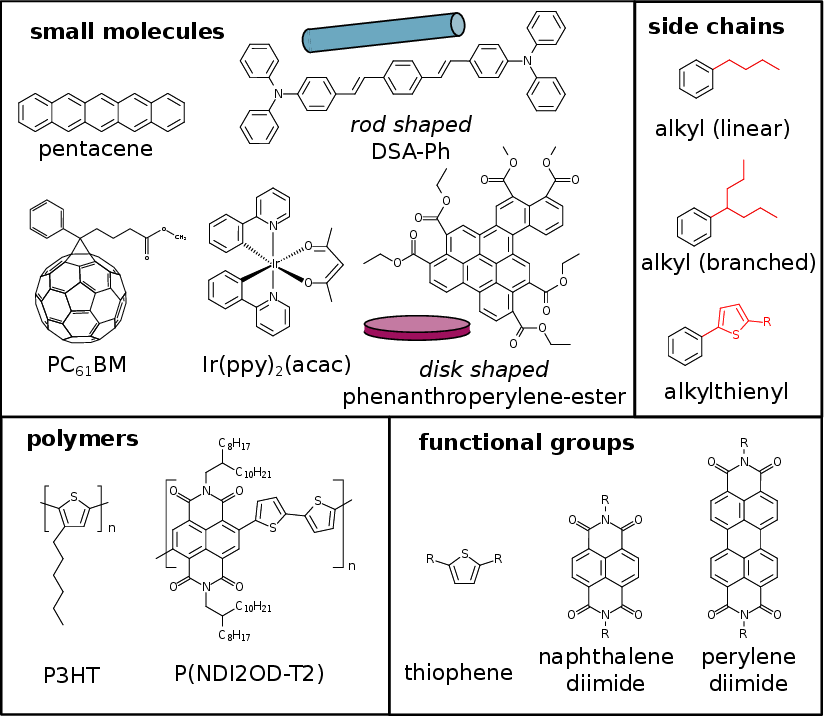}
    \caption{Chemical structures of a number of molecules and functional groups relevant to this work.}
    \label{fig: chemical structures}
\end{figure*}

The chemical structures of molecules and functional groups relevant to this work are shown in figure \ref{fig: chemical structures}.

%%%%%%%%%%%%%%%%%%%%%%%%%%%%%%%%%%%%%%%%%%%%%%%%%%%%%%%%%%%%%%%%%%
%----------------------------------------------
%    THEORETICAL BACKGROUND
%----------------------------------------------

\section{Theoretical Background}
\label{Section: Theoretical background}

The interesting optoelectronic properties of organic semiconductors arise from significant delocalization of $\pi$ electrons due to extended conjugation (alternation of single and double or triple bonds) in the molecular structure, as illustrated in figure~\ref{fig: chemical structures}. Accordingly,  organic semiconductor molecules generally consist of highly anisotropic rigid subunits such as fused aromatic rings. This molecular shape anisotropy has consequences for the interfacial supramolecular assembly of organic semiconductors and optoelectronic processes, as detailed in this and the following sections.

The presence of an interface between a system of anisotropic particles with another phase (e.g. a gas, solid, liquid) breaks the translational symmetry of the system and can induce orientational ordering at the interface, even when the bulk phase is isotropic. For a collection of biaxially anisotropic particles, in which each particle has three inequivalent principal axes, the possible preferred orientations with respect to the interface can be classified into three extremes -- end-on, face-on, or edge-on (figure~\ref{Fig: end, face, edge-on}) -- with the actual orientation for a specific system possibly being intermediate between these extremes. For conjugated molecules such as organic semiconductors, variations in the preferred orientation will change the orientation of the $\pi$-stacking direction with respect to the interface, which has implications for exciton and charge dynamics. 

\begin{figure}[hbt]
\centering
\includegraphics[width = 8 cm]{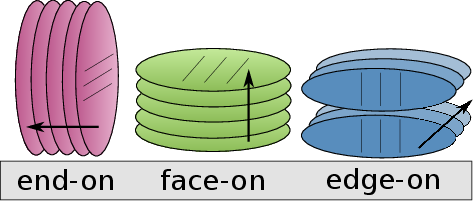}
\caption{Classification of orientations of a collection of biaxially anisotropic particles with respect to a surface. Arrows indicate the direction of the shortest molecular axis, which is generally the $\pi$-stacking direction in organic semiconductors. The surface is shown in grey.}
\label{Fig: end, face, edge-on}
\end{figure}

Orientational ordering of anisotropic particles at interfaces has long been studied, particularly in the domain of liquid crystals \cite{DeGennes1995}, in which experimental, theoretical, and computational studies on this topic have been extensively reviewed \cite{Jerome1991,TeloDaGama1995,Bates1999,Ryu2016}. Most of this work has focused on the equilibrium behavior of single-component fluids of uniaxial mesogens, in which each molecule has one inequivalent and two equivalent principal axes. Thus, the full complexity of organic-semiconductor systems, which often comprise multi-component blends of biaxial molecules whose microstructure is formed under non-equilibrium conditions, cannot be fully understood based on these studies. Nevertheless, this extensive body of work provides a basic conceptual framework for understanding interfacial alignment of organic semiconductors. 

In the field of liquid crystals, orientational ordering with respect to an interface is called surface anchoring \cite{Jerome1991}, and is classified as homeotropic, planar, or tilted, depending on whether the (non-equivalent) axis of the (uniaxial) molecule is perpendicular, parallel, or oblique to the plane of the interface. Early Landau--de Gennes-type phenomenological theories \cite{DeGennes1971}, in which the free energy is expanded in powers of a nematic order parameter, have provided useful insight into surface anchoring, but do not lend themselves to simple interpretation in terms of the microscopic properties of the intermolecular interactions. With more recent work using molecular field theories (e.g. generalised van der Waals theories and classical density functional theories) \cite{TeloDaGama1995} and molecular simulations \cite{Bates1999}, a clearer picture is emerging of the molecular parameters that govern interfacial orientational ordering of liquid crystals.

One general finding of these studies is that the direction of interfacial alignment of liquid crystals is non-universal \cite{Bates1999,MartinDelRio1995}, and depends on the details of the intermolecular interactions, such as the competition between anisotropic repulsive and attractive interactions \cite{Kimura1985}. Using a generalised van der Waals theory, it has even been shown that different terms in a spherical harmonic expansion of the anisotropic attractive interactions of a model mesogen can have opposite effects on interfacial alignment \cite{MartinDelRio1995}, revealing the subtle interplay of factors that control orientational ordering.  Nevertheless, some general trends can be gleaned from theoretical and computational studies of model liquid crystal interfaces.  

Most theoretical and computational studies have focused on the liquid--vapor interface or nematic--isotropic interface between coexisting nematic and isotropic liquid phases of a single-component fluid. For purely repulsive intermolecular interactions, orientational ordering is governed by excluded-volume entropic effects, which result in perpendicular alignment of prolate molecules at the vapor interface \cite{Kimura1985,MartinDelRio1995}. On the other hand, for molecules with both (short-range) repulsive interactions and (longer range) attractive interactions, various alignments are obtained depending on the anisotropy of the molecular shape and interaction strength. Most molecular simulations of liquid crystals with attractive interactions have used the \ac{GB} potential \cite{Gay1981,Bates1999}, which is a general anisotropic potential energy function that depends on the relative position and orientation of pairs of ellipsoidal molecules. For a uniaxial molecule, the molecular shape anisotropy in the \ac{GB} potential is characterized by the parameter $\kappa$, which is the ratio of the molecule length to breadth, while the interaction strength anisotropy is characterized by the parameter $\kappa'$, which is the ratio of the attractive well depth for side-to-side versus end-to-end interactions. For both prolate \cite{Mills1998,MartinDelRio1997,Rull2017} and oblate spheroids \cite{Rull2017} for a range of temperatures and $\kappa$ and $\kappa'$ values, alignment at the vapor interface has been shown to be controlled by the $\kappa/\kappa'$ ratio, with $\kappa/\kappa' > 1$ yielding planar (parallel) alignment and $\kappa/\kappa' < 1$ giving perpendicular alignment. This behavior was rationalized based on the relative energies of different cleavage planes of close-packed ordered arrays of molecules. This essentially amounts to the exposed interface being the one with the lowest interfacial tension, and suggests that interfacial ordering is energetically rather than entropically controlled by the parameters chosen in these systems. 

Compared with the vapor interface, orientational ordering of a liquid crystal at a solid interface is complicated by the influence of the interaction with the solid. Thus, theoretical and computational studies of this situation have been more limited and a general theoretical understanding is lacking \cite{TeloDaGama1995}. 
For purely repulsive hard-core interactions between prolate molecules and with a solid surface, entropic excluded-volume effects favor planar alignment with the surface \cite{Okano1983}, 
opposite to the perpendicular alignment found for similar molecules at the vapor interface \cite{Kimura1985}.  Similarly, semiflexible polymers, which in a sense can be considered highly anisotropic prolate molecules,  confined by a repulsive solid surface tend to align parallel to it \cite{Egorov2016a}. For molecules with both repulsive and attractive interactions with each other and with the solid substrate, the interfacial ordering depends on the details of the substrate--molecule interaction strength and anisotropy \cite{Wall2003}. For example, planar anchoring is observed for strong coupling between model \ac{GB} ellipsoidal particles and a solid substrate at which the substrate--molecule interactions favor this alignment. On the other hand, for a fluid confined between solid and vapor interfaces  that is weakly coupled to a solid,  the alignment at the solid surface is controlled by that at the vapor interface at temperatures below the nematic--isotropic transition \cite{Wall2003}.

\begin{figure}[hbt]
\centering
\includegraphics[width = 8cm]{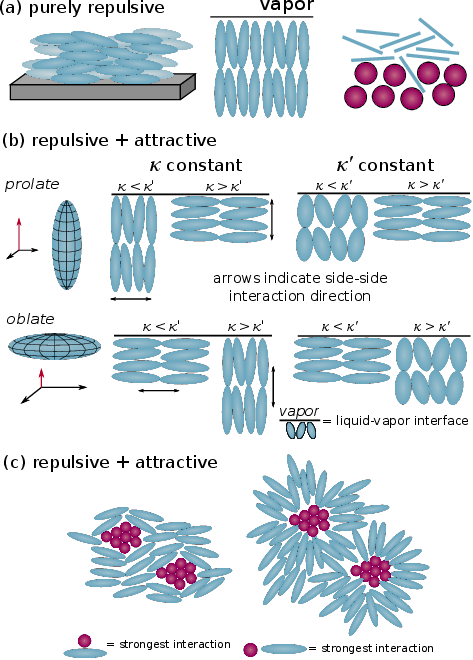}
\caption{(a) Orientation of (left) hard rods at the interface with a solid substrate, or (center) a vapor, and (right) \revision{needles at a fluid interface}. An ellipsoid with purely repulsive interactions has been shown to have a preference for parallel alignment at the solid and fluid interface, and perpendicular at the vapor. (b) Orientation of an ellipsoid, having both attractive and repulsive interactions, with its vapor, and (c) with a spherical fluid. For an ellipsoid at the vapor interface (b), increasing the side--side interaction strength (decreasing $\kappa$', left) or increasing aspect ratio (increasing $\kappa$, right) gives a parallel orientation for both prolate (top) and oblate (bottom) ellipsoids. The red arrow indicates the axis that defines orientation. The direction perpendicular to this defines the side--side direction. At the interface with a fluid (c), orientation depends on the strength of the interactions between either the end or side of the ellipsoid with the spheres.}
\label{Fig: interface ordering}
\end{figure}

For interactions of anisotropic particles at the interface with another fluid, the strength of the interactions between the particles is, again, an important factor for controlling interfacial alignment \cite{Antypov2004}, and so general predictive rules for orientational ordering have to our knowledge not been developed. For purely repulsive interactions between a mixture of hard spheres and vanishingly thin needles, classical density functional theory has shown that the needles  have a slight preference for planar anchoring at the sphere--needle interface in the demixed state \cite{Brader2002}. 
On the other hand, for molecules with both repulsive and attractive interactions, \ac{MD} simulations of a system of model \ac{GB} rods and \ac{LJ} spheres \revision{have shown that} increasing the strength of the interactions between the size of the rod and the spheres gives a preference for planar anchoring, while stronger interactions between the end of the rod and the spheres gives perpendicular alignment \cite{Antypov2004}. These effects are summarized in figure \ref{Fig: interface ordering}.

%%%%%%%%%%%%%%%%%%%%%%%%%%%%%%%%%%%%%%%%%%%%%%%%%%%%%%%%%%%%%%%%%%

%---------------------------------------------------
% SECTION 3: 
% ROLE OF INTERFACES AND SUPRAMOLECULAR ASSEMBLY
%---------------------------------------------------

\section{Role of interfaces and supramolecular assembly}
\label{Section: Role of interfaces}
%-------------------------------------------------------------

It has long been known that molecular anisotropy at interfaces is important for enhancing charge transport and device performance, with \citet{Sirringhaus1999} reporting significantly higher \ac{FET} mobility with edge-on orientations of \ac{P3HT} than face-on orientations of the same molecule almost 20 years ago \cite{Sirringhaus1999}. Since then, many studies of have found similar correlations between orientation and device performance for \ac{OLED} \cite{Yokoyama2011, Kim2014, Komino2014, Lee2017a}, \ac{OPV} \cite{Tumbleston2014, Ye2015, Lee2017}, and \ac{OFET} \cite{Nahid2018, Liu2017, Zhang2011a} devices. In general, in-plane  alignment of the transition dipole moment of emitter molecules with respect to the substrate gives better optical properties in \acp{OLED} \cite{Yokoyama2011}, a face-on alignment at donor--acceptor interfaces appears to be generaly preferred  for good performance in \ac{BHJ} \acp{OSC} \cite{Tumbleston2014, Ye2015, Lee2017}, and edge-on alignment is found in most high-performance \acp{OFET} \cite{Kline2007, Sirringhaus1999}, although face-on structures may still give good mobility in some cases \cite{Rivnay2010, Li2012}. It is therefore important that alignment at these interfaces can be controlled. 

%--------------------------------------------------------------------
% 3.1: Semiconductor-substrate interfaces in organic transistors
%--------------------------------------------------------------------
\subsection{Semiconductor--substrate interfaces in organic transistors}
\label{Section: semiconductor-substrate-interfaces}

In organic transistors, charge transport occurs in an organic semiconductor material that has been deposited onto the surface of a dielectric, with transport restricted to an accumulation layer within a few nm of the semiconductor--dielectric interface \cite{Horowitz1998, Horowitz2004, Dinelli2004, Kiguchi2003}. The microstructure at this interface  therefore has a significant effect on the charge-carrier mobility, which is the key performance metric of \acp{OFET} and also plays an important role in \acp{OLED} and \acp{OPV}. Indeed, a linear correlation between molecular tilt angle at the interface and \ac{OFET} saturation mobility has recently been shown \cite{Nahid2018}. Additionally, a number of studies have shown that, although semicrystalline semiconductors show similar charge mobilities on different dielectric surfaces, the mobility of amorphous semiconductors depends strongly on the properties of the dielectric \cite{Zhao2009, Caironi2011, Li2012, Suemori2008, Jimison2008}. Thus, understanding the structures at the interface, which can differ significantly from that in the bulk \cite{Nahid2017, Duong2012}, is important for improving device performance. Thus, the many studies on bulk morphology are not necessarily relevant for correlating electronic processes and morphology in these devices.

\begin{figure*}[hbt]
\centering
\includegraphics[width = 12cm]{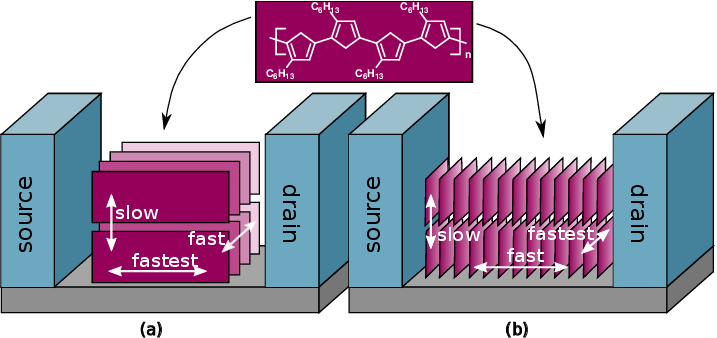}
\caption{Architecture of a bottom-gate bottom-contact \ac{OFET} at the dielectric interface. Polymers are shown in an edge-on conformation with backbone either (a) parallel or (b) perpendicular to the required charge transport direction. Charge transport is fastest along the polymer backbone, slower, but still fast, in the $\pi$-stacking direction, and slowest along the lamellar stacking direction.}
\label{Fig: charge transport speed}
\end{figure*}

The orientational order (edge-on or face-on) and, for polymers, the direction of the backbone with respect to the source and drain electrodes play a significant role  in organic transistor performance. An edge-on alignment to the dielectric surface, where the $\pi$-stacking direction is parallel to the substrate, is expected to enhance charge mobility. This is due to the much higher mobility of charge carriers in the $\pi$-stacking direction, related to the substantial overlap of the $\pi$-orbitals, than through the insulating aliphatic chains (figure~\ref{Fig: charge transport speed}), which would occur in a face-on orientation \cite{Khim2018}. For polymers, a backbone direction parallel to the charge-transport direction (i.e. between source and drain electrodes) will give the fastest charge transport, as charge carriers can move along the backbone (figure \ref{Fig: charge transport speed}) \cite{Khim2018}; however, this will be a function of chain length as shorter polymer chains will require more interchain hops along this direction \cite{Khim2018}.  On the other hand, small molecules give the greatest performance when the $\pi$-stacking direction is the same direction as charge transport, as this facilitates hopping between adjacent small molecules in the desired direction \cite{Pisula2009}. Indeed, when methods that induce order in certain directions are used for the generation of films, highly anisotropic charge transport has been observed \cite{Trefz2019, Bucella2015}.

The effect of liquid-crystalline order in organic semiconductors on charge transport has been comprehensively reviewed \cite{Pisula2009}, from which some general conclusions can be drawn about the influence of molecular anisotropy. Disc-shaped molecules give one-dimensional columnar structures featuring good $\pi$-orbital overlap and high charge transport, but also a large susceptibility to defects \cite{Pisula2009}. Rod-like mesogens on the other hand, give smectic phases that do not pack as closely as the columnar phases formed by discs but are more resilient to defects due to the possibility of two-dimensional charge transport. Finally, board-like polymers, such as polythiophenes, combine both the close $\pi$-stacking of the discs with the two-dimensional charge transport of rods, giving enhanced charge carrier mobility \cite{Pisula2009}. 

A number of factors have been shown to influence edge-on versus face-on orientation with respect to the dielectric interface, ranging from the molecular weight (for polymers) \cite{Nahid2017, Noriega2013, Kline2003}, to the side chain \cite{Zhang2011a, Chen2013a} and backbone \cite{Chen2013, Chen2013a} structure, the substrate composition \cite{Yoneya2017a, Kline2007, Nguyen2015, Bucella2015, Yan2012, Chen2013a}, deposition technique \cite{Trefz2019, Lee2010a, Wurzbach2019, Khim2018}, and the solubility of the material \cite{Chen2013a, Chen2013, Nahid2017, Nahid2018, Luzio2013, Sapolsky2018, Bucella2015}. These methods will be discussed in greater detail in following section in the general context of controlling anisotropy in organic semiconductors. However, as the direction of the backbone of polymer semiconductors with respect to source and drain electrodes is uniquely applicable to organic transistors, this factor will be be examined here.

The solubility of organic polymers has been shown to be an important consideration when considering device morphology as the choice of solvent can either drive or prevent self-aggregation leading to significant changes in directionality, influencing the charge transport. For example, for bar-coated films of the commonly used semiconducting polymer \ac{N2200}, it was found that when a solvent was used in which significant pre-aggregation in the solution occurred, charge transport was highly anisotropic, being significantly greater parallel to the deposition direction than perpendicular to it  \cite{Bucella2015}. The same effect was not observed for the same polymer in a solvent in which the extent of pre-aggregation was lower, with similar mobilities in both directions, indicating a more isotropic film. The mobility in the deposition direction was almost an order of magnitude higher than that for a film prepared by spin-coating, which would be expected to produce more isotropic films. By contrast, in the perpendicular direction  the mobility was significantly lower, emphasizing the importance of this anisotropy for high-performing devices \cite{Bucella2015}. 

For molecules with a permanent dipole moment, preferential alignment of these dipole moments has implications for energetic disorder and thus for charge-carrier mobility \cite{Friederich2017,Friederich2018a}. Energetic disorder can also be affected by interactions with the dielectric, which can be controlled by interfacial orientation. \citet{Richards2008} showed using a simple analytical model that the electronic structure at the semiconductor--dielectric interface can differ significantly from that in the bulk and is significantly influenced by the dielectric constant of the dielectric material \cite{Richards2008}: a higher dielectric constant caused a broadening of the density of states and greater energetic disorder due to the interaction between the randomly oriented dipoles in the gate dielectric and the charges in the semiconductor, leading to a reduction in mobility. However, the effect was shown to decay rapidly with distance (within a few \AA{} from the interface), so could be limited through the introduction of something separating the semiconductor backbone and dielectric, such as a \ac{SAM} of an insulating material. The inclusion of a \ac{SAM} was shown to give a more ordered energetic landscape and improved performance \cite{Sirringhaus2010}. Even a thin layer of alkyl chains, such as the side-chains of an edge-on orienting polymer, has been shown to be sufficient to reduce the broadening of the density of states and effectively remove the dependence of charge mobility on substrate dielectric constant \cite{Zhao2009}, highlighting again the importance of interface orientation on transistor performance.

%--------------------------------------------------------------------
% 3.2: Host-guest/substrate interfaces in OLEDs
%--------------------------------------------------------------------
\subsection{Host--guest/substrate interfaces in organic light-emitting diodes}
\label{Section: interfaces-in-OLEDs}

The active layer of an \ac{OLED} is composed predominantly of a host charge-transporting material, commonly a conjugated small molecule but sometimes a polymer, doped with an emitter material. These materials are vacuum or solution deposited onto a solid substrate to make the device. The orientation of both the emitter and the host with respect to the substrate are therefore relevant, although they are expected to be influenced by similar factors.

The orientation of the emitter molecule in \acp{OLED} has a significant effect on device performance, with in-plane (horizontal) alignment of the molecule's transition dipole moment with respect to the substrate, corresponding to light emission in the perpendicular (vertical) direction, increasing light out-coupling efficiency and external electroluminescence quantum efficiency and thus boosting performance \cite{Yokoyama2011, Kim2014, Komino2014, Lee2017a}. Controlling emitter orientation has thus been the subject of extensive study in recent years. A number of factors have been shown to influence the alignment of emitter molecules with respect to their substrate, including the choice of host \cite{Moon2015, Senes2017}, the shape of the emitter molecule \cite{Senes2017, Youn2018}, strength of interactions with the substrate \cite{Youn2018, Friederich2017, Friederich2018a}, emitter chemical structure \cite{Jurow2016}, the presence of permanent dipole moments \cite{Noguchi2012}, processing technique \cite{Jurow2016, Lampe2016}, and temperature \cite{Komino2014, Youn2018, Dalal2015, Walters2017, Antony2017, Lyubimov2015, Jiang2016, Gujral2017}.

\begin{figure}[hbt]
\centering
\includegraphics[width = 8cm]{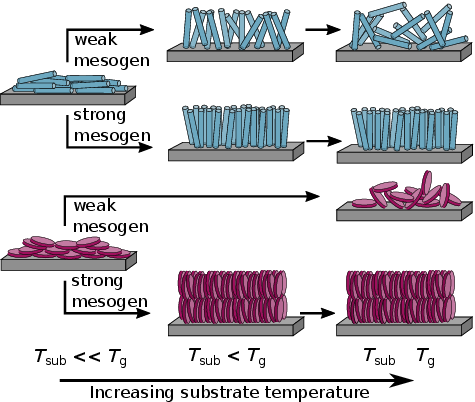}
\caption{Effect of substrate temperature of the orientation of (top) rod- and (bottom) disc-like molecules with both solid and vapor interfaces. While both rod- and disc-shaped molecules are initially deposited in a face-on orientation at the solid substrate, at sufficiently high temperatures reorientation of rod-shaped molecules towards a end-on structure has been observed, before becoming isotropic at \ac{Tsub} $\approx$ \ac{Tg}. Disc-shaped molecules did not show such a marked preference for end-on at intermediate temperatures and formed isotropic glasses at higher temperature. Strong mesogens of both shapes were shown to form robust end-on structures at intermediate temperatures \revision{that are maintained up to }and up to above \ac{Tg}.}
\label{Fig: Disc-mesogen assembly}
\end{figure}

\ac{OLED} devices are often produced by vapor deposition, through which good control of in-plane alignment can be achieved. The active layers of these materials often lack long-range crystalline order, and can be considered glasses. The anisotropy of vapor-deposited organic glasses has been extensively studied and found to be primarily controlled by the ratio of the \ac{Tsub} to the \ac{Tg}. For a series of rod- \cite{Dalal2015} and disc-like \cite{Gujral2017} small molecules of various aspect ratios, at \ac{Tsub}~$\ll$~\ac{Tg} a significant preference for face-on alignment with respect to the substrate was found.  At temperatures just below \ac{Tg}, the rod-like molecules transitioned towards end-on structures, before becoming isotropic at and above \ac{Tg}, although disc-shaped mesogens were found to form robust columnar phases in an end-on orientation even above \ac{Tg} \cite{Gujral2017}. Similarly, rod-like mesogen itraconazole was also found to maintain end-on structure at and above \ac{Tg} \cite{Gomez2016}. Greater orientational anisotropy around \ac{Tg} was also found for longer molecules \cite{Yokoyama2011, Dalal2015, Youn2018}. Additionally, all-atom \ac{MD} simulations of ethylbenzene, a model glassy system which can be considered to have similar structural properties to common organic semiconductors showed the same dependence on \ac{Tsub} \cite{Antony2017}. Furthermore, \ac{CG} \ac{MD} simulations of rod-like or disc-like molecules showed the same trends as found experimentally despite the models only being parameterized to reproduce the molecular shape and not the specific interactions of the experimentally studied molecules \cite{Dalal2015, Walters2017, Lyubimov2015}. This dependence on temperature also agrees with \ac{MD} studies of vapor-deposited glasses of Alq$_\textrm{3}$, a common host material in \acp{OLED} \cite{Bagchi2019},  further atomistic \ac{MD} studies on a series of rod-shaped molecules \cite{Youn2018}, and experimental studies of a similar, but shorter, rod-shaped molecule in a randomly oriented host \cite{Komino2014}. It should be noted that the aforementioned studies predominantly examined single-component systems, while the active layer in \acp{OLED} contains an emitter molecule doped into a host matrix, generally at low concentrations. However similar trends have been observed for two-component systems. \citet{Jiang2016} showed, through experimental studies of the blue light emitter DSA-Ph, a rod-like small molecule, in Alq$_\textrm{3}$, a similar dependence on the \ac{Tsub}:\ac{Tg} ratio, which was effectively independent of concentration. The \ac{Tsub} dependence described is outlined in figure \ref{Fig: Disc-mesogen assembly}.

To explain these trends, deposition has been proposed to proceed by a `surface equilibration mechanism'. The free surface of the deposited glass has been shown to have significantly higher mobility than the bulk so that as deposition occurs, for sufficiently high substrate temperatures, molecules near the surface are able to reorient towards the equilibrium orientation for this interface \cite{Ediger2019}. Interestingly, for the range of molecules studied, this preferred orientation was end-on, which is consistent with the orientation expected for mesogens with purely repulsive intermolecular interactions at the vapor interface (section~\ref{Section: Theoretical background}), although the attractive interactions present in real systems may mean that this behaviour is not universal. At low substrate temperatures, the deposition of additional layers kinetically traps the layers below in the face-on orientation that molecules most likely adopt when they first collide with the surface. This effect, as well as other factors that effect orientation in anisotropic organic glasses, has been recently reviewed \cite{Ediger2019, Gujral2018} and the reader is referred to these for further details.

Although mechanisms for achieving good control over parallel alignment are known for vapor-deposition processing methods, films fabricated through solution processing techniques have been shown to give generally isotropic films \cite{Jurow2016,Lampe2016}. As solution processability is one of the reasons organic semiconductors are so attractive, controlling alignment through these methods is of general interest. \citet{Senes2017} recently investigated some methods for controlling the orientation of the transition dipole moment of a rod-like molecule in solution-processed \acp{OLED}, finding that better parallel alignment of the transition dipole moment, which was aligned with the long axis of the molecule, could be achieved through the use of a semiflexible polymeric host molecule that showed preferential alignment in the plane of the substrate, as expected from the discussion in section~\ref{Section: Theoretical background}.  Thermal annealing of the system was shown to further increase anisotropy to levels comparable with vacuum-deposited layers. Additionally, through comparison of a number of different emitter molecules with different backbone shapes, they found that molecules with elongated, more rod-like, backbones more readily formed parallel aligned structures, an effect which has also been observed for vapor-deposited films \cite{Youn2018, Dalal2015, Yokoyama2011}.

%------------------------------------------------------------------------
% 3.3: Bulk-heterojunction donor-acceptor interfaces in organic photovoltaics
%------------------------------------------------------------------------
\subsection{Bulk-heterojunction donor--acceptor interfaces in organic photovoltaics}
\label{Section: BHJ-donor-acceptor-interfaces}

The photoactive layer of \ac{OPV} devices generally consists of a blend of an electron donor and an electron acceptor material called a \ac{BHJ} that is phase separated on the nanoscale \cite{Jiao2017}.  Charge separation and  recombination at the donor--acceptor interface is a major factor limiting device efficiency \cite{Menke2018a}. The microstructure, including molecular orientation, at electron donor--acceptor interfaces in \ac{BHJ} \acp{OPV} plays a potentially important, but as yet unclear, role in determining good device performance. Charge generation at donor--acceptor interfaces is known to be more efficient than predicted by classical electrostatic models at a structureless interface \cite{Deibel2009}, suggesting that the inhomogeneity of the interface may play a role in enhancing charge separation. Indeed, changes in orientation have been implicated in increasing the efficiency of charge generation \cite{Ran2017}, improving efficiency of separation and charge transport away from the interface \cite{Poelking2015}, and reducing exciton recombination during charge separation \cite{Tumbleston2014, Ye2015}. 

A number of possible mechanisms for enhanced charge separation and reduced recombination have been proposed (which potentially all play a role) that depend on the interfacial microstructure. For example, the electron--hole binding free energy at the interface has been shown theoretically to be reduced by delocalization of charges \cite{Deibel2009} or by increased energetic disorder \cite{Hood2016}, which would be expected to be enhanced and diminished, respectively, by increased structural order at the interface. Structural variations at the interfaces could also induce electronic-energy gradients that promote charge separation \cite{Jamieson2012}. Computational studies have also shown that the electric field due to aligned molecular quadrupoles at the interface can drive charge separation \cite{Poelking2015}, with the direction of the field and thus the spatial variation of the interfacial electronic energies sensitive to the molecular orientation at the interface \cite{Verlaak2009,Davino2016}. Furthermore, interfacial alignment will affect the electronic coupling between donor and acceptor. It has been suggested that a face-on orientation at the interface is generally preferred, as it increases orbital overlap between donor and acceptor \cite{Tumbleston2014, Ye2015}. Additionally, the energy of interfacial \acp{CTS} is very sensitive to the arrangement of donor and acceptor at the interface \cite{Baumeier2012};  for a pentacene:C$_\textrm{60}$ interface, a face-on orientation has been shown computationally to give a higher energy \ac{CTS}, attributed largely to a change in the electron affinity of the C$_\textrm{60}$ acceptor due to the induced electrical field, which is expected to reduce non-radiative recombination and voltage loss in an \ac{OPV} device \cite{Chen2016}.  

With the myriad potential roles of molecular orientation at \ac{OPV} heterojunction interfaces, controlling orientation at these interfaces is an important step towards better predictability and reproducibility of device properties. Furthermore, the ability to measure or predict the interfacial molecular orientation is crucial to clarifying the relative importance of the various charge-separation mechanisms discussed above. Measurements of orientational order at donor--acceptor interfaces have only recently become possible with the development of polarized soft X-ray scattering (P-SoXS) techniques \cite{Jiao2017}, and so data relating interface orientation to molecular structure or processing conditions remains limited. As described in section~\ref{Section: Theoretical background}, controlling intermolecular interactions between materials at the interface is important for directing orientation. Increasing the planarity of the molecule is one means to do this, as it would increase the extent of the interaction between donor and acceptor. This has been achieved by adding fluorine atoms along the conjugated polymer backbone, which has been shown to give a preference for a face-on orientation and better device performance \cite{Tumbleston2014}. In this case, intramolecular interaction between appropriately positioned fluorine and backbone sulfur atoms restrict backbone rotation, increasing planarity of the conjugated backbone \cite{Kim2013}. Alternatively, increasing the extent of conjugation, such as by the substitution of alkyl side chains for conjugated alkylthienyl ones, would likewise be expected to enhance interactions between molecules, and has indeed been shown to increase the extent of face-on orientation and enhance performance for a mixture of polymer-donor and non-fullerene acceptor \cite{Lee2017, Ye2015}.

%%%%%%%%%%%%%%%%%%%%%%%%%%%%%%%%%%%%%%%%%%%%%%%%%%%%%%%%%%%%%%%%%%

%---------------------------------------------------
% SECTION 4: 
% GENERAL PRINCIPLES FOR CONTROLLING ORIENTATION
%---------------------------------------------------

\section{Towards general principles for controlling interfacial orientation}
\label{Section: General principles}

As noted in the previous section, the orientation of semiconductors with respect to their interfaces is an important property for improving performance in a variety of organic-electronic device types, from \ac{BHJ} organic solar cells to \acp{OLED} and organic transistors. In this section, we examine the factors that have been shown to influence the interfacial orientation of organic semiconductors and attempt to unify these observations into some general rules for controlling molecular orientation. These factors will be broadly classified into material properties, and thermodynamic and processing conditions. While \citet{Osaka2015} have recently reviewed structural and processing methods for controlling backbone orientation in semiconducting polymers, we aim to extend this more broadly to non-polymeric organic semiconductors and to other factors that have been shown to influence anisotropy in both in-plane and out-of-plane directions at interfaces. While not an exhaustive review of the literature of control of interface anisotropy, we have endeavored to identify the key properties that to influence interface orientation based on research to date.

%----------------------------------------------------
% material properties
%----------------------------------------------------
\subsection{Material properties}
The properties of the molecules themselves (semiconductors, substrates, and solvents) play an important role in controlling the orientation of semiconductors at interfaces. Through modification of the molecular shape, both energetic and entropic driving forces can be modulated to give preference to different orientations, while intermolecular interactions can also directly influence the alignment properties of semiconductors.

\subsubsection{Shape anisotropy}
\paragraph{Rod-shaped molecules: aspect ratio}
As explained in section~\ref{Section: Theoretical background}, hard prolate particles, with purely repulsive interactions, preferentially align perpendicular to a vapor interface \cite{MartinDelRio1995, Kimura1985}, and parallel to a solid interface \cite{Okano1983} due to excluded-volume entropic effects (figure~\ref{Fig: interface ordering}). This type of behaviour has been observed experimentally in a number of real systems for a range of molecules of different shapes and lengths. Studies of vapor deposition, in which molecules were deposited from the gas phase onto a solid substrate, have shown that rods of various lengths prefer to orient perpendicularly to the vapor interface following deposition \cite{Dalal2015, Youn2018, Gujral2017, Ediger2019}. In this case, as they were deposited, the molecules oriented parallel to the substrate, but at higher temperatures the surface molecules (at the vapor interface) reoriented towards an end-on (perpendicular) orientation, and, if mesogenic, formed strongly end-on liquid-crystalline phases \cite{Gujral2017, Gomez2016}. This mechanism is discussed in greater detail in section~\ref{Section: interfaces-in-OLEDs}. The orientation preferences described are not unexpected, having been shown to be the preferred orientation for purely repulsive prolate particles. Given that the influence of attractive interactions are expected to effect orientational preference \cite{MartinDelRio1997, Rull2017}, it is interesting to note that all of the molecules studied exhibited the same alignment behavior.  Coarse-grained molecular simulations of particles parameterized just to reproduce molecular shape have shown similar trends to both experimental and all-atom simulations \cite{Walters2017, Dalal2015}, indicating that molecular shape is an important parameter for determining orientation at the vapor interface. 

In addition, longer molecules have been shown, through computationally studies, to show stronger orientational anisotropy than their shorter counterparts \cite{Dalal2015, Youn2018}. We note that these studies did not distinguish between face-on and edge-on orientations as they generally used models with uniaxial symmetry, but it seems reasonable that a similar argument can be applied to this additional orientational anisotropy and that high aspect-ratio molecules will initially orient with a face-on preference in vapor deposition.

\paragraph{Rods vs discs}
As for molecules with other shapes, vapor-deposited disc-shaped molecules of varying size have also been shown, through \ac{MD} simulation, to orient predominantly face-on to the substrate at low temperatures \cite{Walters2017, Gujral2017}. At high temperatures, again, orientation becomes isotropic, but for disks that are \revision{strong} mesogens, a transition to an edge-on columnar stacking at \ac{Tsub} close to and above \ac{Tg} has been observed \cite{Gujral2017}. As with the rods, this is consistent with the reorientation of ordered structures when free to rotate to give the equilibrium end-on structure.

\paragraph{Backbone length and planarity}
As described above, increasing the length of the molecular backbone has been shown to give more extreme orientational preference, whether parallel or end-on. Alternatively, for polymers, the planarity of the backbone has also been shown to play a role in interfacial orientation. \citet{Chen2013} found that while coplanar backbones can give either face-on or edge-on orientation depending on the solubility of the compound, slight deviations from coplanarity predominantly gives edge-on orientation. 

\subsubsection{Intermolecular interactions}
As discussed in section~\ref{Section: Theoretical background}, changes to non-bonded (attractive) intermolecular interactions for fixed molecular-shape anisotropy can lead to dramatic changes to surface anchoring (e.g. from planar to perpendicular). Thus, these interactions are expected to strongly influence molecular orientation at organic semiconductor interfaces. Strong attraction between the face of an organic semiconductor and a solid or another molecule would give a face-on orientation, whereas if the interactions between the face of the conjugated molecule and the substrate were unfavorable, or the interactions of the substrate with the molecules edges stronger, an edge-on orientation would be preferred. This is consistent with reports from molecular simulation of a semiconducting polymer at different solid interfaces where low substrate--semiconductor interaction strength gives edge-on oriented polymer and high interaction strength gives face-on \cite{Yoneya2017a}. In addition to influencing the orientation at the interface, the strength of the interaction can also influence the distance from the interface to which the orientational order is maintained,  with stronger \ac{vdw} interactions between substrate and semiconductor have been shown to give a thicker oriented interface \cite{Youn2018}. 

\revision{In the organic-semiconductor field, intermolecular interactions at interfaces are often characterized in terms of surface energy. Correlations between substrate surface energy and interfacial alignment have recently been observed for semiconducting polymer films, with lower surface energy associated a more edge-on interfacial orientation than a higher energy surface \cite{Zhang2018a}. This is consistent with favourable interactions between the polymer alkyl side chains with the similar sidechains of a low-energy \ac{SAM}-treated surface, which is maximized in the edge-on orientation. Similar correlations have been observed in the organic single-crystal literature, with crystal orientation and grain morphology showing a strong dependence on substrate properties \cite{DonPark2007}, and orientated growth being achieved through adjustment of surface energy and epitaxial growth due to similar lattice parameters \cite{Zhang2018h, Liu2009b}. In line with the results of semicrystalline polymers, crystalline pentacene, for example, has been shown to adopt a perpendicular orientation with respect to a low surface energy substrate, as this exposes its lowest energy face to the substrate, while adopting a face-on interaction with high-energy substrates where its higher energy face interacts with the substrate \cite{DonPark2007}.}

As described in section~\ref{Section: Theoretical background} for molecules at a vapor interface, increasing end--end interaction strength has been shown to give greater parallel alignment. Practically, one means of achieving this is through the use of rod-like molecules that are able to form weak hydrogen bonding networks end-to-end with each other. These stronger end--end interactions have been shown to give more horizontally aligned vapor deposited films than otherwise very similar molecules without the hydrogen bonding capability \cite{Watanabe2019}. A number of other means of tuning interaction strength between semiconductor and substrate exist, including factors such as changing the chemical composition of the backbone, side chain modification, and changes in molecular shape, which will be discussed in detail in the subsequent sections.

\begin{figure}[hbt]
\centering
\includegraphics[width = 6 cm]{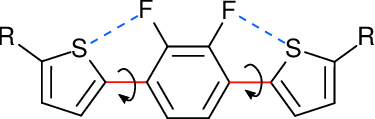}
\caption{Restriction of backbone rotation due to F--S interactions (dashed blue). The red bonds, indicated with an arrow, are less free to rotate in the fluorinated molecule.}
\label{Fig: fluorine rotation}
\end{figure}

\paragraph{Atomic substitution: fluorination}
Atomic substitution, such as the substitution of hydrogen for the highly electronegative fluorine is a powerful means for tuning interaction strength. As fluorine is fairly small, the substitution can be made without adversely impacting steric constraints. The addition of intramolecular non-bonded interactions between backbone fluorine and sulfur atoms has been shown, for example, to be a useful method for inducing liquid-crystalline order in solution-processed conjugated polymers as it induces backbone planarization at high concentration, promoting the formation of aggregates displaying liquid crystal-like behaviour \cite{Kim2013} (figure~\ref{Fig: fluorine rotation}). At donor--acceptor interfaces in polymer--fullerene solar cells, face-on orientation of fluorinated polymer at the interface has been observed, while the non-fluorinated equivalent was slightly edge-on \cite{Tumbleston2014, Osaka2015}.

\paragraph{Side chains}
Side chains for organic semiconductors are predominantly engineered to alter the solubility and thus influence the aggregation properties of the molecule in solution, which can be directly tuned by changing side chain density, length, or connectivity. The effect of aggregation on orientation will be discussed later in the context of solubility, but we note that side chain arrangements that are likely to prevent aggregation, such as bulky and densely distributed alkyl chains, have been shown to give face-on morphologies for conjugated polymers at a solid surface, while those that promote ordered packing, such as linear, sparsely distributed chains are more likely to give edge-on structures \cite{Osaka2015, Chen2013a}. Additionally, \citet{Osaka2014} reported that difference in side chain length for a polymer with two different side chains on each monomer can strongly impact orientation at the substrate interface. Side chains that were approximately the same length were found to give face-on structures, while if the side chains were significantly different in length an edge-on orientation was preferred \cite{Osaka2014}. Finally, polymer regioregularity, where the monomers are all coupled in the same head-to-tail fashion in the polymer chain, should enhance the formation of ordered aggregates which may also influence anisotropy \cite{Kim2006}.

While factors such as side chain length and density affect packing through changes to steric interactions, changing the side chains to modulate the strength of interactions at a  semiconductor interface is also an interesting method to influence orientation. For example, stronger interactions could be introduced between molecules at the interface by replacing pure alkyl side chains with alkylthienyl chains to give 2D conjugated structures. This has been shown to give a more face-on orientation at the donor--acceptor interface in all-polymer solar cells \cite{Lee2017, Ye2015}.

\subsubsection{Polymer specific factors: molecular weight and chain stiffness}
Polymer \ac{MW} is known to have a substantial effect on device performance \cite{Kang2015, Osaka2012, Kline2003, Zhou2016, Nahid2017, Fan2017, Bartelt2014} with higher \ac{MW} polymers generally (although not always) giving better performance. Although not applicable to small-molecule semiconductors, polymer \ac{MW} is known to affect whether the molecule orients edge-on or face-on with respect to its substrate, which may at least in part explain its effect on device performance. \citet{Osaka2015} previously surveyed the literature for the effect of \ac{MW} on backbone orientation and found that in general, high \ac{MW} polymers form edge-on morphologies with respect to their interfaces, while low \ac{MW} generally gives randomly oriented polymers. It should be noted, however, that this depends on the polymer itself, and \revision{may not be} completely generalizable for all polymers. For example, more recently,  \citet{Nahid2018} found that for semiconducting polymer \ac{N2200}, the presence of rod-like aggregates formed by low \ac{MW} polymers in poor solvents \cite{Nahid2017} gave a more edge-on orientation at the polymer--air interface and better mobility than their higher \ac{MW} counterparts. However, they also noted that although decreasing \ac{MW} gave a more edge-on orientation in the bulk, there was little difference at the dielectric interface, which remained predominantly edge-on over all \acp{MW}. Nevertheless, this indicates that different classes of polymer may behave differently with respect to \ac{MW}, and other factors may need to be considered, in particular the shape and size of aggregates formed, which will be discussed in greater detail below. 

In their \ac{MD} and theoretical studies of generic conjugated polymers at substrate interfaces, \citet{Zhang2016} found that increasing \ac{MW} increased the distance from the substrate at which polymers were still aligned. Although isotropic in the bulk, when coming in contact with the surface, the polymers aligned parallel to the surface, to a distance approximately equal to the polymer's persistence length for polymers longer than this length.  For longer and stiffer chains, this ordered layer was found to be thicker, due to a larger nematic coupling parameter, with longer molecules retaining some order even slightly beyond the persistence length. They found that a higher \ac{MW} was associated with a greater preference for the parallel orientation. However, as their simulations considered uniaxial molecules, they were unable to distinguish between edge-on and face-on orientations of the polymer.  

%----------------------------------------------------
% processing conditions
%----------------------------------------------------
\subsection{Thermodynamic and processing conditions}

\subsubsection{Temperature}
Temperature is an important property for controlling interface orientation as it controls both the equilibrium microstructure and the rate at which the structure evolves towards equilibrium. Considering the example of vapor deposition, \ac{Tsub} significantly lower than semiconductor \ac{Tg} has been shown to give parallel alignment of a range of disk- and rod-shaped molecules, while an increase in temperature to just below \ac{Tg} shifts the alignment of rods towards perpendicular. A further increase in temperature beyond \ac{Tg} leads to isotropic phases \cite{Dalal2015, Walters2017, Antony2017, Youn2018, Lyubimov2015, Komino2014, Jiang2016, Gujral2017, Bagchi2019}. As discussed previously, this is attributed to a surface equilibration mechanism in which only the top couple layers of the deposited film are sufficiently mobile to allow realignment of molecules \cite{Ediger2019}. The molecules are initially deposited in the face-on orientation and then have only a finite amount of time to realign before they are buried. At low temperatures, the molecules are kinetically trapped in the as-deposited face-on orientation. As the temperature is increased, the molecules can reorient towards the equilibrium orientation, which appears to be edge-on for a range of small-molecule organic semiconductors, before their orientation is trapped by the deposition of further layers on top. Above \ac{Tg} interactions between molecules are generally not strong enough to maintain order and anisotropy is lost. Although this phenomenon may also be explained by a change in the thermodynamically favoured morphology with temperature as the balance of entropy to enthalpy changes, studies of deposition rate found \cite{Ediger2019} at lower deposition rates a favored edge-on orientation was reached at lower \ac{Tsub}, which suggests that the ability to reorganize before becoming trapped under additional layers is the driving factor for the observed temperature dependence.

Temperature in other processing methods may also be expected to influence semiconductor alignment for similar reasons. As devices are often fabricated under non-equilibrium conditions, the orientation they are initially deposited at (either from vapor or solution) can potentially become kinetically trapped. For semiconducting polymer \ac{N2200}, annealing of a solution processed system above its melting point has been shown to change the orientation at the top surface and in the bulk from face-on to edge-on \cite{Rivnay2011}, while blade coating at similar temperatures has been shown to induce a greater degree of edge-on polymer \cite{Trefz2019}. For P3HT at the donor-acceptor interface with a fullerene, interfacial orientation has similarly been shown to shift towards edge-on with annealing \cite{Bhattacharyya2019, Osaka2015}. For these cases, it is interesting to note that in all cases of which we are aware  the edge-on orientation appears to be the equilibrium structure, but as surface anchoring orientation is not expected to be universal, this may not be favored in general. Further investigation into the equilibrium structures of anisotropic particles (both polymeric and single molecule) at interfaces is necessary to better understand this. 

\subsubsection{Solvent}
As solution processing methods are common in the production of organic-electronic devices, solvent is a simple parameter to tune, and is known to have a significant effect on polymer conformation. It has been proposed that high mobility in polymer semiconductors can be achieved using poor solvents that induce a greater degree of pre-aggregation (aggregation in solution prior to deposition on the substrate), as these can give enhanced liquid-crystalline ordering \cite{Luzio2013}, which is generally associated with more regular alignment. Even for non-liquid-crystalline materials, the formation of large aggregates has been observed to give rise to edge-on orientations at the substrate interface \cite{Chen2013a, Trefz2019}. The larger the aggregate, the greater the effect and the stronger the preference for edge-on structures \cite{Trefz2019}. This is consistent with recent reports of correlation between the extent of aggregation of \ac{N2200} in solution and edge-on orientation at the film--air interface \cite{Nahid2018}. However, previous reports have observed the opposite trend for the same polymer, with pre-aggregation leading to a predominantly face-on orientation, though not specifically at the interface \cite{Steyrleuthner2014}. These opposing trends can be reconciled with the knowledge that this polymer has been shown to preferentially display edge-on orientation at the film--air interface, and face-on orientation relative to the air interface in the bulk \cite{Schuettfort2013}, suggesting that the orientation at the substrate surface may be templated by the orientation at the vapor interface or in the bulk depending on the deposition conditions.

As with temperature, although the molecules examined seem to consistently favour edge-on orientations when aggregates are formed, there is no obvious physical principle that describes this dependency, nor is it clear whether an edge-on orientation for aggregates is preferred more generally. Again, studies employing generic models may be useful for elucidating the effects of aggregate formation on interfacial orientation. As solubility can be tuned by a number of factors, such as side-chain modifications or choice of solvent and solvent additives, this provides a relatively simple way of controlling interface orientation, so a more thorough understanding of how it affects interfacial ordering would be beneficial.

\subsubsection{Processing techniques: external forces}

Processing techniques are well known to influence the anisotropy of molecular interfaces with techniques that introduce shear or flow forces, such as blade or bar coating, giving more anisotropic films than, for example, spin coating \cite{Jurow2016, Lampe2016, Trefz2019, Wurzbach2019}. These non-equilibrium methods rely on the introduction of orientation-specific forces to promote different morphologies and alignment directions. Methods such as blade coating are able to give extended alignment of both fiber direction and orientation of polymeric materials, with significant alignment of chains in the shear direction observed \cite{Wurzbach2019, Trefz2019, Khim2018}. Similarly, the application of an external magnetic field has been shown to give alignment of the backbone of a conjugated polymer \cite{Pan2015} or discotic mesogen \cite{Shklyarevskiy2005} parallel to the field due to anisotropy in the molecule's diamagnetic susceptibility. \revision{Likewise, alignment of P3HT fibres in the direction of an applied electric field has been observed, giving increased charge mobility \cite{Mas-Torrent2004}.} Light has also been shown to be able to specifically align liquid crystals \cite{Simoni1999} although has not been studied explicitly in relation to organic semiconductors. A number of other methods for aligning and orienting polymer films by means of processing conditions and techniques have been recently reviewed \cite{Khim2018} and the reader is referred to that work for greater detail.

\revision{Additionally, these processing techniques have been extensively used for crystalline organic semiconductors, with mechanical stretching, external magnetic- or electric-fields and solution deposition methods such as zone casting all having been shown to give good control over orientation \cite{Liu2009b}. Alternatively, concentration and temperature gradients, tilted substrates, and solution shearing have all also been shown to allow for control of crystallization direction \cite{Li2016g}. Modulation of the shear rate, for example, has been shown to change the packing density of small molecule TIPS-pentacene, with the optimal rate giving a closer $\pi$--$\pi$ stacking distance and enhanced charge carrier mobility \cite{Li2016g}.}

%----------------------------------------------------
% summary
%----------------------------------------------------
\subsection{Summary}
From the preceding discussion, it should be clear that, although there are many factors which have been shown to influence interfacial anisotropy of organic semiconductors, many of them have not yet been systematically studied. Although some interesting trends are starting to develop further study is required if general principles for controlling molecular alignment at semiconductor interfaces are desired. While compelling mechanisms for some effects have been proposed, such as the surface equilibration mechanism for vapor deposited films, others, such as the effect of solution aggregation or processing temperature on orientation cannot yet be described universally through general physical principles. Likewise, although orientation of uniaxial repulsive particles at the vapor interface, and to a lesser extent solid and fluid interfaces, is fairly well understood, the role of attractive interactions, which are important for real materials, and the balance between repulsive and attractive interactions, is less clear. A better understanding of these interactions may enhance our understanding of the preferential alignment observed in relation to modifications such as aggregation, fluorination, and backbone structure. 

%%%%%%%%%%%%%%%%%%%%%%%%%%%%%%%%%%%%%%%%%%%%%%%%%%%%%%%%%%%%%%%%%%

%---------------------------------------------------
% SECTION 5: 
% MODELLING ORGANIC SEMICONDUCTOR INTERFACES
%---------------------------------------------------

\section{Modeling organic semiconductor interfaces}
\label{Section: Modeling interfaces}

As the microstructure at interfaces in organic semiconductor devices is known to be important for device performance, being able to characterize the interfacial microstructure and its relationship to electronic processes would facilitate the design of better devices. Although possible, it is often experimentally challenging to characterize these interfaces as they are generally buried within the device \cite{Salleo2007, Salleo2010}. Additionally, it is difficult, if not impossible, to extract molecular-level detail from experimental data of organic semiconductor films due to their significant disorder. With the ability to directly simulate and visualize these interfaces on an atomic scale, computer simulations are an attractive method for uncovering the intricacies of structure and assembly mechanisms of organic semiconductor interfaces. This section reviews applications of computer simulations to elucidate the microstructure and assembly of organic semiconductor interfaces. Since most of the computer-simulation techniques discussed in this section and their general use to study organic-semiconductor morphology has previously been comprehensively reviewed previously \cite{Dantanarayana2012,Muccioli2014,Gartner2019}, the techniques themselves will only briefly be described here, with the focus being on applications that clarify the role of interface anisotropy. 

We will also not go into detail on methods for simulating the electronic structure and electronic processes at organic semiconductor interfaces, as these methods have been extensively reviewed in the past \cite{Beljonne2011,Few2014,Davino2016}. Since fully quantum-mechanical simulations that account for both nuclear and electronic degrees of freedom are unfeasible for simulating microstructure formation, a general computational approach that has been widely adopted for investigating the microstructure dependence of electronic processes in organic semiconductors is to use computationally efficient classical methods that do not explicitly account for electronic degrees of freedom to simulate the physical structure and structural evolution and then to apply a quantum-mechanical or mixed quantum--classical approach  \cite{Beljonne2011,Few2014,Davino2016} to the obtained physical structure to calculate electronic properties. This approach is expected to be reasonable, as the structural dynamics responsible for microstructure formation are generally not expected to be strongly influenced by the electronic processes in these systems.  Thus, this section will focus only on the first step of this approach: the simulation of the physical structure of organic-semiconductor interfaces. 

%------------------------------------------------------
% all-atom simulation
%------------------------------------------------------
\subsection{All-atom simulations}

A number of different methods can be applied to simulate molecular systems and provide geometries that can be used for detailed electronic calculations. Broadly, these can be classified as \ac{MD} or \ac{MC} simulations. \ac{MD} is particularly useful for the study of organic semiconductor interfaces as it allows direct study of the system dynamics and can therefore capture nonequilibrium processes that can be important in organic-semiconductor interface formation. On the other hand, \ac{MC}  algorithms generally sample configurations from an equilibrium statistical ensemble and therefore are most adapted to simulating equilibrium states. Nevertheless, hybrid \ac{GC} \ac{MC}--\ac{MD} simulations, which involve \ac{MC} particle insertion and deletion steps, can be effective for studying nonequilibrium interface formation in the presence of processes such as solvent evaporation. For further details on these methods, the reader is referred to \cite{Frenkel2001} and \cite{Allen1987}. 

A distinction should be made between atomistic (all-atom) and \ac{CG} particle-based simulations. In atomistic simulations, every atom is treated explicitly, whereas \ac{CG} models group atoms together into a single interaction site, decreasing the degrees of freedom of the system at the expense of atomic resolution with the goal of reducing computational expense. Both methods will be discussed here.

%-----------------------------------------------------

\subsubsection{Background of molecular dynamics}
Classical \ac{MD} is a molecular-simulation technique that in essence integrates Newton's classical equations of motion to study the evolution of a molecular system over time \cite{Allen1987,Frenkel2001}. Interactions between particles are defined through a force field which is used to give the total potential energy as the sum of all bonded and non-bonded interactions in a system. A number of standard force fields exist which are generally transferable between a large number of molecules. However, these force fields often need to be modified in order to accurately model the intermonomer dihedral-angle potentials in polymers, which are strongly affected by the $\pi$ conjugation inherent to organic semiconductors \cite{DuBay2012, Jackson2015}.

%-----------------------------------------------------

\subsubsection{Study of physical structure and assembly}
\begin{figure*}[htb]
\centering
\includegraphics{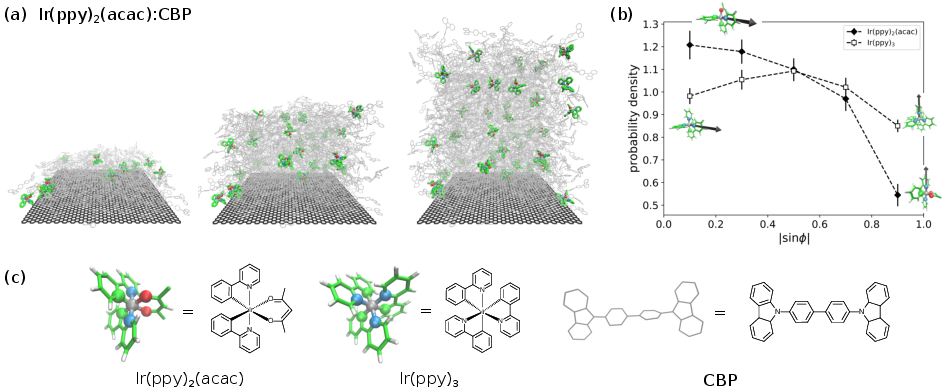}
\caption{\revision{(a) Snapshots of a simulation of vapor deposition of a host--emitter OLED system onto a graphene substrate (black, thick line representation) \cite{Lee2017a}. The host is CBP (grey lines) and the emitter is Ir(ppy)$_\textrm{2}$(acac) (green). Hydrogens are omitted for clarity. (b) Transition dipole moment (TDM) orientation for Ir(ppy)$_\textrm{2}$(acac) (filled diamonds) and Ir(ppy)$_\textrm{3}$ (unfilled squares) from deposition simulations \cite{Lee2017a}. A value of $\vert\sin \phi\vert$ of 0 corresponds to horizontal alignment of the transition dipole moment and 1 to vertical, as shown in the overlaid structures. The TDMs for both molecules are assumed to lie along the Ir--N bonds. In the structures, only one TDM is shown for clarity as a black arrow.  (c) Chemical structures of Ir(ppy)$_\textrm{2}$(acac), Ir(ppy)$_\textrm{3}$, and CPB. For the iridium complexes: green = carbon, blue = nitrogen, red = oxygen, grey = iridium, white = hydrogen. The molecules in (a) are colored accordingly. }}
\label{Fig: deposition Lee2017a}

\end{figure*}

The use of atomistic \ac{MD} simulation to study the physical structure and assembly at organic semiconductor interfaces is becoming more common as computational power increases. In particular, the process of vapor deposition has been extensively studied by \ac{MD} for a variety of substrates and small molecules \cite{Hu2018a, Tonnele2017, Ratcliff2015,Yoo2019, Lee2017a, Youn2018, Han2015, Muccioli2011, Walters2017, Antony2017} (figure \ref{Fig: deposition Lee2017a}). While these simulations generally agree well with experimentally observed phenomena, such as the dependence of orientation on \ac{Tsub}, they also provide further insight into the mechanisms of the process. \citet{Muccioli2011}, for example, simulated vapor deposition of pentacene onto a C$_\textrm{60}$ substrate. As pentacene molecules were added, they found that they initially lay flat on the surface, before diffusing rapidly and beginning to aggregate. As the aggregates on the surface grew, they coalesced into a film covering the surface with random horizontal alignment. As more pentacenes are added to the first monolayer, the molecules rearranged to form an end-on morphology with a herringbone pattern. Additional layers then follow a similar mechanism \cite{Muccioli2011}.

In a similar vein, the deposition of host--emitter systems onto a substrate, relevant to \acp{OLED}, has been studied using atomistic \ac{MD} simulation \cite{Tonnele2017,Lee2017a}. Emitter molecules with different degrees of molecular anistropy,  Ir(ppy)$_\textrm{3}$ and Ir(ppy)$_\textrm{2}$(acac), were found to have a preferred orientation with respect to a graphene substrate, resulting in anisotropy in the alignment of the transition dipole moment for phosphescent emission (obtained from quantum-chemical calculations of isolated emitters) that was consistent with experimental light-outcoupling measurements \cite{Lee2017a} \revision{(figure~\ref{Fig: deposition Lee2017a})}. The alignment of the highly symmetric Ir(ppy)$_\textrm{3}$ in particular was proposed to be driven by alignment of the anisotropic host material 4,4'-bis(\textit{N}-carbazolyl)biphenyl (CBP) \cite{Lee2017a}.

\ac{MD} simulations have also recently been used to study the preferred orientation of a  semiconducting oligmer in solution at various interfaces. \revision{At the solution vapor interface, the orientation of the oligomer was found to be edge-on, while when sandwiched between two solid substrates weak semiconductor-substrate interactions gave edge-on orientation and strong interactions face-on. When between a solid substrate and a free vapor interface, edge-on orientation was observed at the vapor interface and either edge- or face-on at the solid surface depending on the substrate} \cite{Yoneya2017a}. From these simulations it was concluded that if interaction between the $\pi$-conjugated plane (face) of the molecules was more energetically favourable than the interactions of the same plane with the interface (either gas phase or solid), the edge-on configuration was preferred. On the other hand, face-on orientation with respect to the substrate was preferred if interactions with substrate surface were stronger than interactions between the conjugated faces of the molecules \cite{Yoneya2017a}.

In \acp{OPV}, the size of \ac{BHJ} domains is generally too large to be feasibly simulated atomisitically, being on the order of 10~nm \cite{Lyons2012}. However, simulations of model systems representative of sections of these interfaces have provided valuable insight. For example, calculations of the \ac{Voc}, based on quantum-chemical calculations of electronic energies using configurations obtained from short \ac{MD} simulations of planar interfaces of a number of different molecules in specific orientations at a C$_\textrm{60}$ surface, have given excellent agreement with experimental values; this allowed the authors to rationalize the high performance of a particularly high-performing series of organic semiconductor through the favourable energetic effects induced by its interfacial orientation \cite{Poelking2015}. Elsewhere, \ac{MD} simulations have shown that at a pentacene--C$_\textrm{60}$ interface, C$_\textrm{60}$ molecules are able to burrow into the surface of face-on, but not end-on, oriented pentacenes \cite{Fu2013}, giving a more disordered region and the type of interphase structures shown elsewhere to lead to enhanced charge separation \cite{Poelking2015a}.

\subsubsection{Challenges for all-atom simulation of interfaces}
Despite these successes, there are still a number of challenges associated with all-atom simulations of organic semiconductor interfaces. Potentially the largest of these challenges is accessing the relevant length and time scales for modeling processes in organic semiconductor materials, which can contain many millions (or more) of atoms and involve processes occurring on up to the second scale, or even longer for annealing processes. Although simulations of up to 100 million atoms (for 100 ns) \cite{Sener2016}, or up to the millisecond timescale for smaller systems using specialized hardware \cite{Shaw2010}, are possible, typical simulations can only study a couple million atoms on the nanosecond (or microsecond at best) timescale, meaning that slower processes occurring on longer timescales are not able to be explicitly studied. This is particularly relevant when considering the scale of structural variations at interfaces, which generally involve large systems whose assembly occurs over large time periods, which can be unfeasible to simulate atomistically. All-atom simulations of polymers have also generally been limited to chain lengths of 10s of monomers, which is 1--2 orders of magnitude smaller than those studied experimentally in organic semiconductor systems, limiting the ability to realistically capture polymer microstructure using such models. 

In order to deal with the problems of size and time scale, while still maintaining atomistic detail, a number of approaches may be taken, potentially at a cost to quantitative accuracy. \citet{Wang2016}, in their study of a polymer-fullerene system representative of a \ac{BHJ} donor--acceptor interface, point out that a phase-segregated morphology cannot be fully achieved with the system sizes available to \ac{MD}. They instead used a higher fullerene concentration than would be used experimentally to be able to qualitatively understand the behaviour at the interface \cite{Wang2016}. Alternatively, \citet{Yoo2019} proposed the `frozen-bulk' method for studies of interfaces in vapor-deposition simulations. They proposed that, at large distances from the interface and at a temperature lower than \ac{Tg}, the orientation of the bulk region does not significantly change. The motion of these regions was therefore frozen to allow for more efficient simulation \cite{Yoo2019}. \citet{Ratcliff2015} also used positional fixing to enhance computational efficiency for the \ac{MC} simulation of vapor deposition. In this work, once the molecules were deposited their positions were fixed. However, it has since been shown that there should still be quite significant movement of the molecules after deposition so this method may not accurately capture all the details of the deposition process \cite{Tonnele2017}. \citet{Bagchi2019} looked at the problem slightly more abstractly, and, as the surface structure of  vapor-deposited glasses has been shown to be very similar to that of the equilibrium liquid \cite{Dalal2015, Ediger2019}, simulated this equilibrium liquid surface, which is likely to display much faster dynamics than the glass, instead.

%------------------------------------------------------
% coarse-grained simulation
%------------------------------------------------------
\subsection{Coarse-grained molecular simulations}

\begin{figure}[hbt]
	\centering
	\includegraphics[width = 8cm]{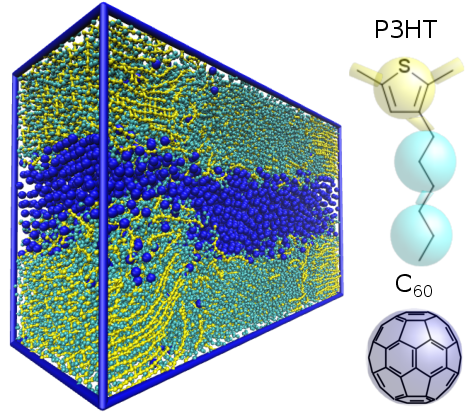}
	\caption{\ac{BHJ} P3HT:C$_\textrm{60}$ donor--acceptor interface from \ac{CG} \ac{MD} simulation \cite{Huang2010}.}
	\label{Fig: CG BHJ}
\end{figure}

While atomistic simulations provide details of specific interactions at interfaces, and give an atomic understanding of the processes occurring at these interfaces, they are often limited by the size of the system and time scales of the processes of interest. A common way to address these problems is the use of \ac{CG} molecular simulations, in which the number of degrees of freedom in the system is greatly reduced by approximating a collection of atoms as a single interacting site, increasing the simulation efficiency of the calculation. It is especially important that larger length scales than are feasible atomistically to be reached when one considers, for example, the molecular weight dependence of interface morphology of polymer semiconductors \cite{Kline2006, Kline2003, Noriega2013, Nahid2017} or the domain sizes in \acp{BHJ} which are typically on the order of around 10~nm \cite{Lyons2012}. In contrast to all-atom simulations, \ac{CG} simulations have been shown to be able to reach appropriate length scales with enough accuracy to \revision{model aggregation of conjugated polymer chains in solution \cite{Schwarz2013} and phase separation in conjugated polymer \acp{BHJ} \cite{Zhou2016, Huang2010, Alessandri2017}} (see figure~\ref{Fig: CG BHJ}). 

%------------------------------------------------------

\subsubsection{Background of \ac{CG MD} simulations} 
\begin{figure}[hbt]
\centering
\includegraphics[width = 8cm]{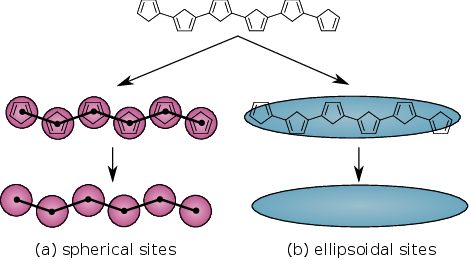}
\caption{An example of coarse graining of small conjugated molecule sexithiophene into (a) six spherical sites, or (b) a single ellipsoidal site. The spherical model represents each thiophene unit as a sphere connected by their centers of mass (black lines to black dots), while the ellipsoidal model is a representation of how this molecule could be coarse-grained into a single anisotropic particle.}
\label{Fig: CGed sexithiophene}
\end{figure}

The process of coarse-graining is summarized schematically in figure~\ref{Fig: CGed sexithiophene}. Groups of atoms expected to have correlated motion are grouped into larger sites whose interactions are parametrized to (hopefully) capture the behavior of the real all-atom system. The total potential energy of the system is calculated as a sum of bonded and non-bonded interactions between \ac{CG} sites. Two general approaches have been used to parameterize \ac{CG} interactions: the top-down approach, in which interactions are chosen to reproduce experimental thermodynamic data \cite{Alessandri2017}, or the bottom-up approach, in which the interactions are tuned to reproduce the physical and thermodynamic properties of an all-atom model \cite{Noid2013}. A number of systematic bottom-up \ac{CG} methods have been developed, with the goal being to achieve thermodynamic consistency between the \ac{CG} and all-atom models, e.g. by matching forces \cite{Izvekov2005} or structural distribution functions\cite{Reith2003} or minimizing the relative entropy \cite{Shell2008} between the two models.  Readers are referred to \cite{Noid2013} for a comprehensive review of coarse-graining methods. 

\subsubsection{Spherical vs anisotropic sites and potentials}
To date, \ac{CG} models have predominantly employed spherical sites for the calculation of non-bonded pair interactions. Although these models have been able to reproduce experimental anisotropic behaviour when the overall molecule (an accumulation of isotropic \ac{CG} sites) is anisotropic (figure \ref{Fig: CG BHJ}) \cite{Schwarz2013, Dalal2015, Lyubimov2015, Zhang2016, Walters2017}, many organic semiconductors consist of large planar subunits with very rigid backbones, so representing them as a collection of spheres may not be accurate or efficient. This is especially problematic when considering properties such as $\pi$-stacking distance, which are known to be important for device performance. An alternative is to use anisotropic, either ellipsoidal \cite{Lee2012, Lee2016} or disc-shaped \cite{Bowen2018}, particles. As these sites now have a quantifiable orientation, anisotropic non-bonded potentials are required to model the inter-site interactions as a function not only of distance, but also of orientation. Although these anisotropic interactions are slower to calculate than isotropic ones, a reduction in the number of sites (figure~\ref{Fig: CGed sexithiophene}) for a given level of accuracy can compensate for this. Interpretation of simulation results of an anisotropic model with fewer degrees of freedom and parameters is also potentially simpler.  A number of anisotropic non-bonded potentials exist that account for both the distance and orientation dependence of the potential energy when considering ellipsoidal sites. Most commonly used are the \ac{GB} \cite{Gay1981, Berardi1995}, and RE-squared \cite{Everaers2003, Babadi2006, Babadi2006a} potentials, which are effective for ellipsoidal particles. Further details can be found in the literature \cite{Babadi2006, Babadi2006a, Everaers2003, Brown2009}.

An alternative potential, more suited to the flat disc-like molecules common for organic semiconductors, has been recently reported \cite{Bowen2018}. Although not a new potential, being originally published over 40 years ago \cite{Stone1978}, the S-function expansion has only recently been applied to organic semiconductors, and with good success \cite{Bowen2018}. Again, details of this potential can be found elsewhere \cite{Bowen2018, Stone1978} so will not be discussed here, other than to note that it may better describe the interactions between these disk-like particles than either the \ac{GB} or RE-squared potentials. 

%------------------------------------------------------
\subsubsection{Study of physical structure and assembly}

\ac{CG MD} models can give valuable insights into mechanistic details of microstructure and assembly processes in organic semiconducting device that cannot be easily studied experimentally and may be computationally inaccessible by all-atom models. \ac{CG} models have only been applied in the last decade or so to studying organic-semiconductor structure and so, compared with all-atom simulations, \ac{CG} simulations of organic-semiconductor interfaces remain quite limited.

Simple \ac{CG} simulations, using general models with interactions that were not parameterized to match any specific system, but rather just to reproduce the shape of the molecules of interest, have helped to provide a molecular understanding of the dependence of orientation at the interface on \ac{Tsub} for vapor-deposited glasses \cite{Dalal2015, Walters2017, Lyubimov2015}. \citet{Lyubimov2015} showed, through the use of a rod-like model of small-molecule semiconductor \ac{TPD}, represented as six spherical sites connected by springs to maintain the desired shape, similar behaviour as observed experimentally for these systems (horizontal orientation at low \ac{Tsub} and a slight vertical preference at \ac{Tsub} just below \ac{Tg}), and were able to propose the surface equilibration mechanism described in section~\ref{Section: interfaces-in-OLEDs}. Significantly, this  and related work \cite{Walters2017,Dalal2015} showed that even without explicit parameterization for atomistic systems, and not including the effects of dipole moments or polarizability, \ac{CG} models that approximately represent molecular shape and interaction strength are able to reproduce anisotropic structural characteristics consistent with experiments.

Using a generic \ac{CG} \ac{MD} model, \citet{Zhang2016} simulated a semiflexible polymer melt at a disordered, impenetrable interface to examine the effect of nematic coupling, influenced by chain length and stiffness, on surface-induced alignment. They showed that the polymer preferentially orients parallel to the surface, forming an aligned layer of about a persistence length thick. For longer and stiffer chains, the thickness of the aligned layer is increased due to stronger nematic coupling. From this, it is predicted that the conjugated polymer \ac{P3HT} can form an alignment layer of approximately 4.5~nm thick, with stiffer polymers likely forming even thicker layers. This sort of orientational alignment is especially relevent for charge transport in \acp{OFET}, for which order near the interface is more important than that in the bulk.

Both bottom-up \cite{Huang2010,Lee2011,Lee2012a,Lin2012b,Jankowski2013,To2014,Lee2016,Long2017a} and top-down \cite{Alessandri2017} \ac{CG} \ac{MD} models have been used to simulate the microstructure and formation of \ac{BHJ} donor--acceptor interfaces for both small molecules \cite{Lee2016,Long2017a} and polymers \cite{Huang2010,Lee2011,Lee2012a,Lin2012b,Jankowski2013,To2014,Alessandri2017}, a process that would be very challenging to study using an all-atom model. Even using a \ac{CG} model, realistically simulating \ac{BHJ} formation is difficult, and most of these studies simulated donor--acceptor interface formation in the liquid phase at elevated temperatures rather than simulating the process of solvent evaporation normally involved in \ac{BHJ} formation. Even when solvent evaporation has been considered \cite{Alessandri2017,Lee2016}, solvent evaporation rates have been many times those in experiments. 

Nevertheless, these simulations provide useful molecular-level insight into the donor--acceptor interface, although only a couple of these studies have addressed the issue of interfacial molecular orientation. In particular, \ac{CG} simulations of model P3HT:PCBM polymer:fullerene interfaces, in which the polymer chains were oriented edge-on, face-on, or end-on to the interface, or in which the polymer chains were amorphous, indicated that the interfacial energy was lower for the ordered configurations than in the amorphous one and that the energy was lowest at the face-on interface \cite{To2014}. This suggested that ordering was favored at the interface compared with the bulk, which was indeed observed, and that the face-on interface was the most stable \cite{To2014}. These results have implications for \acp{OPV}, as the face-on configuration is widely believed to optimize charge separation. On the other hand, \citet{Lee2016} used an ellipsoidal-site \ac{CG} model of the anisotropic small-molecule semiconductor \ac{SMDPPEH} blended with \ac{PC61BM} in the solvent chlorobenzene, the latter two represented by \ac{CG} spheres, to simulate the effects on the nanomorphology of solvent evaporation and shear forces mimicking the process of blade-coating. They showed that increasing the shear rate led to more stacking of the \ac{SMDPPEH}, which is likely to promote hole transport, but also results in larger isolated \ac{PC61BM} domains, hindering exciton dissociation. Thus, they concluded that an optimal shear rate exists that balances these two characteristics, and gives the greatest charge transport \cite{Lee2016}.

%------------------------------------------------------
\subsubsection{Challenges for coarse-grained simulation of organic semiconductors}

As with atomistic \ac{MD} simulations, \ac{CG} simulations are not without their challenges. In particular, the loss of molecular detail can be problematic as details about specific interactions, which  may be important for understanding device properties, may be lost in the \ac{CG}ing procedure. It is therefore important that the desire for efficiency (fewer sites) be balanced with the need for accuracy. For good predictions, the model must be able to capture important intra- and intermolecular rearranglements which may necessitate a greater number of sites in order to prevent information loss \cite{Carbone2008, Root2016}.

Another desirable feature of a \ac{CG} model, which is not necessarily easy to achieve, is transferability to thermodynamic conditions beyond which it was parameterized. Again, a model that is able to capture the molecular rearrangements that occur at changing tempteratures is important \cite{Carbone2008}. Although this is generally likely to require a greater number of sites, transferability may also be improved with the use of anisotropic sites which may retain the relevant degrees of freedom and better describe rotations of planar molecules which cannot be captured with spherical sites. Indeed, \citet{Bowen2018} noted that their anisotropic disc model for fused ring \acp{PDI} was likely to be more transferable than other \ac{CG} models due to the significant amount of detail that is able to be maintained by the anisotropic sites \cite{Bowen2018}. This was due to the sections that were \ac{CG}ed into discs being rigid moieties that generally did not show drastic rearrangements. Accordingly, little information was lost in the \ac{CG}ing procedure and the model was expected to behave more like its atomistic counterpart.

A further degree of transferability that can be challenging in \ac{CG} models is that of a force field for a specific molecule to other molecules in the same class, which maybe differ by just a side chain. For example, the three-site \ac{P3HT} model of \citet{Huang2010} shows good accuracy for the structural properties of the molecule for which it was parameterized, but on extension to other \acp{P3AT} with longer alkyl chains it has been shown that this model is no longer able to reproduce experimental properties \cite{Root2016}. Slight adjustments to this model, by changing the strength of the interactions between side-chain sites and those between backbone sites, were able to give qualitative agreement with experimental properties, but were still unable to achieve quantitative accuracy \cite{Root2016}. \citet{Root2016} noted that, again, the anisotropic shape of the thiophene ring may be important for accurately predicting structural properties and a spherical site model may not be sufficient.

Additionally, despite their increase in accessible time scale relative to atomistic \ac{MD}, \ac{CG} models are still unable to achieve the time scales relevant to processes such as solvent-based film deposition. Although as computing power increases these time scales should begin to become more accessible, alternative methods, such as the continuum methods discussed below or hybrid approaches using \ac{CG} semiconductors and a continuum or implicit solvent will likely be necessary to realistically model such processes. However, care must be taken to account for hydrodynamic interactions (particularly for polymers) \cite{Dunweg2008} and the concentration dependence of the \ac{CG} interactions when solvent degrees of freedom are integrated out.   

%------------------------------------------------------
%    CONTINUUM SIMULATIONS
%------------------------------------------------------
\subsection{Continuum simulations}

Continuum or field-based models of fluid structure, such as classical density functional theories \cite{Wu2007} or statistical field theories (including self-consistent field theory (SCFT)) \cite{Fredrickson2002} describe the fluid in terms of a spatially varying probability density field instead of by the coordinates of discrete particles, in contrast to the all-atom and coarse-grained particle models described in the preceding sections. In practice, such continuum models are solved numerically on a discrete lattice. They offer distinct advantages over particle-based models, in particular in being able to access much longer length and time scales, which is especially relevant for modeling long polymer chains at high densities, for which relaxation times can be prohibitively long even for simulations of \ac{CG} particle-based models.

Nevertheless, applications of continuum models to studying organic-semiconductor microstructure, let alone the ordering of organic semiconudctors at interfaces, have been very limited. This is in part due to the (arguably) greater theoretical complexity of these models compared with particle-based models and the lack of freely available, general-purpose software for solving such models numerically \cite{Arora2016}, with no such software existing to our knowledge that accounts for orientational degrees of freedom. Furthermore, these continuum theories generally assume equilibrium conditions, with the solution to the model being the set of field variables (e.g. the probability density of a particular molecule type at a position and orientation) that minimizes a system free energy that is a functional of the fields. Although dynamic variants of continuum models have been developed \cite{Wu2007,Fredrickson2002}, they generally assume time evolution of the field variables on the equilibrium free energy surface, which is not obviously correct far from equilibrium. This may limit applicability for modeling the microstructure of the significant proportion of organic-semiconductor systems formed under non-equilibrium conditions.

Most continuum simulations of organic semiconductors (e.g \cite{Buxton2006a,Xue2012}) have used models in which the field variables depend only on position and not on orientation, and so are incapable of modeling orientational ordering. Only a handful of studies\cite{Shah2010,Zhang2015,Zhang2016} have accounted for molecular orientation. 

Zhang et al.~\cite{Zhang2016} studied the orientational ordering of semiflexible polymer chains at an impenetrable surface using an SCFT model of Gaussian chains in which the chain configuration depended on both the position and orientation with respect to the surface. The model was parameterized to represent chains of the conjugated polymer P3HT. Consistent with previous findings for semiflexible polymers \cite{Egorov2016a}, the surface was found to induce alignment of the chains parallel to the surface, with the thickness of the aligned layer on the order of the persistence length. Nevertheless, since the model did not account for monomer orientation, it was not able to distinguish edge-on versus face-on alignment of interest in electronic applications.

Shah and Ganesan~\cite{Shah2010} used SCFT to model the self-assembled bulk-heterojunction morphologies of donor--acceptor rod--coil block copolymers between two surfaces representing solar-cell electrodes. Their model included a Flory--Huggins term for the rod--coil repuslive interaction, a Gaussian stretching energy for the coil blocks, and the anisotropic Maier--Saupe potential for orientational ordering of the rod blocks. The solar-cell device properties were simulated using a classical drift--diffusion model of charge and exciton dynamics that accounted for the effect of orientational anisotropy through a hole mobility that was a function of the orientational order parameter in the rod block obtained from the SCFT model. Using these simulations, the effect of substrate--polymer interactions, rod--coil miscibility, and the degree of orientation order on solar-cell performance was studied. Nevertheless, the consideration of only block copolymers, which allows microphase separation to be observed at equilibrium, constrains the orientational order at the donor--acceptor interface by chain connectivity; so the model is of limited utility for understanding orientational ordering at interfaces between donors and acceptors on different molecules, which are used in most bulk-heterojunction devices.

%----------------------------------------
%    SUMMARY AND OUTLOOK
%----------------------------------------

\section{Summary and outlook}

As highlighted in this review, the structure at interfaces in organic semiconductor based devices is known to be important for device function with, for example, in-plane alignment with respect to the substrate of the transition dipole moment in \acp{OLED} giving better optical properties, face-on alignment at donor--acceptor interfaces in \ac{BHJ} \acp{OSC}, and edge-on alignment at the dielectric interface in \acp{OFET} generally associated with increased performance. It is therefore important to develop an understanding of the potential ways to control interface structure for the realization of efficient, commercializable, organic electronics. Although many specific examples exist of orientational ordering of organic semiconductors at interfaces and its consequences for electronic processes and device properties, a general understanding of the factors that control interfacial alignment in these systems remains lacking, even for equilibrium systems. While the alignment of uniaxial nematics at the vapor interface has been widely studied \cite{MartinDelRio1995, Kimura1985, MartinDelRio1997, Rull2017}, their alignment at solid and fluid interfaces is less well understood, particularly when both repulsive and attractive intermolecular interactions are present. For biaxial nematics, which are more representative of organic semiconductors, the literature is even more sparse with few, if any, studies on their general alignment at either solid, fluid, or vapor interfaces.

In this work we have attempted to collate the many observations, from both experimental and computational work, about factors that affect interfacial microstructure and reconcile these, where possible, with general physical principles to develop some general guidelines for achieving orientational control. In general, it can be concluded that

\begin{enumerate}
\item vapor deposition of small molecules is controlled by substrate temperature, with higher substrate temperatures giving a slight preference to end-on orientation at the air interface, consistent with studies of purely repulsive anisotropic particles at the vapor interface (although the effect that strong attractive interactions in a real system would have on this alignment is unclear),
\item annealing of solution-processed polymer films appears to give preference to edge-on structure at the solid interface in many cases, although it is unclear whether this is universal and further study is required to explain this observation,
\item the strength of attractive interactions between the side or end of the molecule with other molecules and a solid or fluid can control orientation,
\item aggregation of polymers in solution prior to deposition appears to generally favor edge-on orientations while single molecules favor face-on, although the reasons for this, and whether it is is a general property beyond the systems studied, is unclear,
\item and, for semiconductors that display liquid-crystalline properties, the use of external forces, such as shear or magnetic fields, is a potentially powerful method for controlling orientation.
\end{enumerate}

We note that in many cases the studies used to determine these properties generally only focus on a specific polymer or small molecule. Further studies using simple generic models of semiconductor-like molecules, in addition to systematic experimental studies, would be of great use in better understanding the interplay between the properties described above, and developing a general framework for deliberate and precise control of interfacial orientation. \revision{For \ac{BHJ} \acp{OPV} in particular, very little experimental data is available on the structure at donor--acceptor interfaces. With the advent of polarized soft X-ray scattering techniques that allow for experimental characterization of buried interfaces, systematic experimental studies would be of great value in both verifying computational models and developing greater understanding and accurate predictions of interface structure.} Computationally, the use of \ac{CG} models, parameterized to model the typical range of interactions in organic semiconductors, would be useful to extend the known surface-anchoring effects of equilibrium systems of uniaxial molecules to biaxial molecules (where face-, end-, and edge-on orientations can be distinguished), nonequilibrium conditions, and the examination of the interplay of attractive and repulsive interactions. \revision{Particularly in cases in which trends are beginning to emerge but for which universality has not yet been confirmed, such as in the influence of solution aggregation or temperature on interfacial orientation, these systematic studies have the potential to be of great worth. A more thorough understanding of the physical principles underlying the observed orientational preferences, coupled with knowledge of how these structural changes correlate with changes in electronic properties and device efficiency, is therefore the next logical step towards improving the performance of organic semiconductors. The ability to predict solid-state microstructure and electronic properties and processes from the molecular structure of the component materials promises to facilitate a first-principles approach to the design of high-performance organic semiconductors with application in next-generation electronic devices.}

Finally, the use of computer simulations to study organic semiconductor-substrate interfaces is a very powerful method to correlate structural anisotropy with electronic processes, as well as directly visualizing molecular structure in the regions of interest. There are a number of challenges associated with both atomistic (accessing the relevant length and time scales), and \ac{CG} (transferability, loss of information) simulations, but these are rapidly being addressed. The use of anisotropic \ac{CG} sites is of particular interest as these should better be able represent the important structural features of the predominantly disk-like molecules prominent in the organic semiconductor literature and enable the significantly longer simulations required to capture the second-scale (or longer) processes important for self-assembly at interfaces. Further development of continuum models that can describe orientation order or hybrid \ac{CG}--continuum models could be particularly fruitful for realistically modelling the nonequilibrium deposition processes involved in the formation of  organic-semiconductor interfaces.

%%%%%%%%%%%%%%%%%%%%%%%%%%%%%%%%%%%%%%%%%%%%%%%%%%%%%%%%%%%%%%%%

\ack{This work was supported by the Australian Research Council under the \textit{Discovery Projects} funding scheme (DP190102100). BJB acknowledges The University of Adelaide for the Joyner and Constance Fraser scholarships and the Playford Memorial Trust for a PhD scholarship. HTLN thanks The University of Adelaide for a Master of Philosophy (No Honours) International Scholarship.}

%%%%%%%%%%%%%%%%%%%%%%%%%%%%%%%%%%%%%%%%%%%%%%%%%%%%%%%%%%

\newcommand{\newblock}{}
%\printbibliography
\bibliographystyle{unsrtnat}
\bibliography{anisotropy_review}

\end{document}